\documentclass[fleqn,twoside]{article}
\usepackage{espcrc2}

\usepackage{graphicx}
\usepackage[figuresright]{rotating}

\newcommand{\AmS}{{\protect\the\textfont2
  A\kern-.1667em\lower.5ex\hbox{M}\kern-.125emS}}
\newcommand{\lsim}{\mathrel{\mathop{\kern 0pt \rlap
  {\raise.2ex\hbox{$<$}}}
  \lower.9ex\hbox{\kern-.190em $\sim$}}}
\newcommand{\gsim}{\mathrel{\mathop{\kern 0pt \rlap
  {\raise.2ex\hbox{$>$}}}
  \lower.9ex\hbox{\kern-.190em $\sim$}}}

\title{Cold Dark Matter and Neutralinos}

\author{N. Fornengo\address[UniTO]{Department of Theoretical Physics,
 University of Torino \\ and Istituto Nazionale di Fisica Nucleare,
 Sezione di Torino, \\ via P. Giuria 1, I--10125 Torino, Italy \\
 Email: {\sl fornengo@to.infn.it} \\ Web: {\sl www.to.infn.it/$\,\tilde{~}$fornengo/} and
 {\sl www.to.infn.it/astopart/}} \thanks{Based on work done in
 collaboration with A. Bottino and S. Scopel and on a collaboration
 with A. Riotto and S. Scopel on low--reheating cosmological models.}
 }
       
\begin{document}

\begin{abstract}
Neutralinos are natural candidates for cold dark matter in many
realizations of supersymmetry. We briefly review our recent results in
the evaluation of neutralino relic abundance and direct detection
rates in a class of supergravity models.
\vspace{1pc}
\end{abstract}

\maketitle

\section{Introduction}

The fact that most of the Universe is dark and composed of new, exotic
components has been established over the years by means of various
sets of observations, which recently have converged into a consistent
picture where the Universe is about critical and dominated by an
exotic form of matter and by an unexpected type of dark energy. In
terms of the density parameter $\Omega$, the current view can be
summarized as follows: the total amount of matter/energy of the
Universe is $\Omega_{\rm tot} \simeq 1$ at the 10\% level and this is
composed of a matter component $\Omega_{\rm M} \simeq 0.3$ and a
vacuum--energy component $\Omega_{\Lambda} \simeq 0.7$
\cite{TAUP01_Proc}. The existence of both dark exotic matter and dark
energy asks for extension of the standard model of fundamental
interactions, since no known particle or field can represent either of
these components. In this paper, we will deal with the problem of
explaining the observed amount of dark matter, which we can summarize
as: $0.05 \lsim \Omega_{\rm M} h^2 \lsim 0.3$, and with the studies
related to the searches for dark matter particles. For an updated
review on these subjects, see Ref. \cite{TAUP01}.

\section{Neutralino dark matter in Supergravity}

Supersymmetric models with $R$--parity conservation naturally predict
the existence of a stable relic particle. The nature and the
properties of this particle depend on the way supersymmetry is
broken. In particular, the neutralino happens to be the dark matter
candidate in models where supersymmetry is broken through gravity--
(or anomaly--) mediated mechanisms. The actual implementation of a
specific susy scheme depends on a number of assumptions on the
structure of the model and on the relations among its parameters. This
induces a large variability on the phenomenology of neutralino dark
matter.

The simplest and most direct implementation of supersymmetry is
represented by the {\em minimal supergravity} (mSUGRA) scheme, where
gauge coupling constants are unified at the GUT scale and at the same
scale also the susy--breaking mass parameters are universal. The
low--energy sector is obtained through renormalization group evolution
of all the parameters of the model, and this also induces the breaking
of the electroweak symmetry in a radiative way. This model is very
predictive, since it relies only on four free parameters. However,
neutralino phenomenology is quite constrained \cite{TAUP01,noi,others}
and also quite sensitive to some Standard Model parameters, like the
mass of the top and bottom quarks ($m_t$ and $m_b$) and the strong
coupling constant $\alpha_s$ \cite{noi}. Less constrained susy
implementations are obtained by relaxing the universality conditions
({\em non--universal SUGRA models} \cite{noi}) or by defining the
relevant susy framework directly at the electro--weak scale as an
effective low--energy theory ({\em low--energy minimal supersymmetric
standard model} \cite{noi}).

\begin{figure}[t] \centering
\vspace{-20pt}
  \includegraphics[width=19pc]{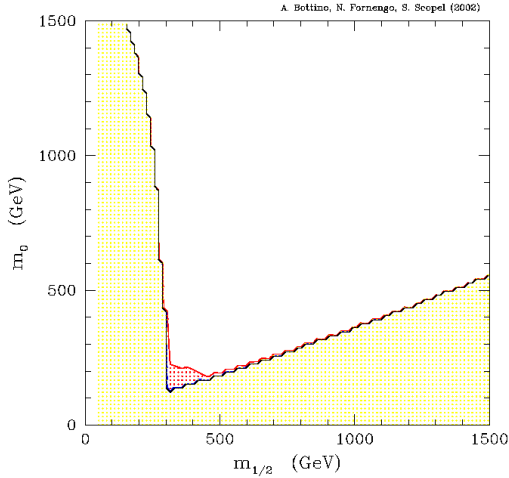}
\vspace{-50pt}
\caption{$m_{1/2}$--$m_0$ plane at $\tan\beta = 30$ in mSUGRA. The
dotted (red) region denotes where the neutralino relic abundance lies
in the cosmologically relevant range $0.05 \leq \Omega_{\rm M} h^2
\leq 0.3$. The light--dotted (yellow) area is excluded by experimental
and theoretical constraints.
\label{fig:sugra_relic_tb30}}
\end{figure}

\begin{figure}[t] \centering
\vspace{-20pt}
  \includegraphics[width=19pc]{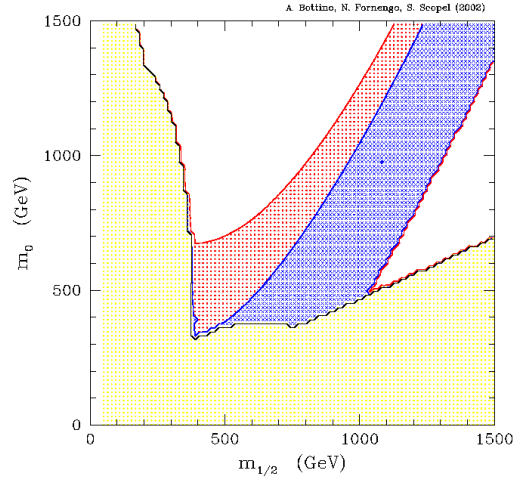}
\vspace{-50pt}
\caption{The same as in Fig. 1, for $\tan\beta = 50$. The dotted (red)
light regions denote where the neutralino relic abundance lies in the
cosmologically relevant range. The hatched (blue) region refers to a
cosmologically sub--dominant neutralino, {\em i.e.}  $\Omega_\chi h^2
< 0.05$.
\label{fig:sugra_relic_tb50}}
\end{figure}

\begin{figure}[t] \centering
\vspace{-20pt}
  \includegraphics[width=19pc]{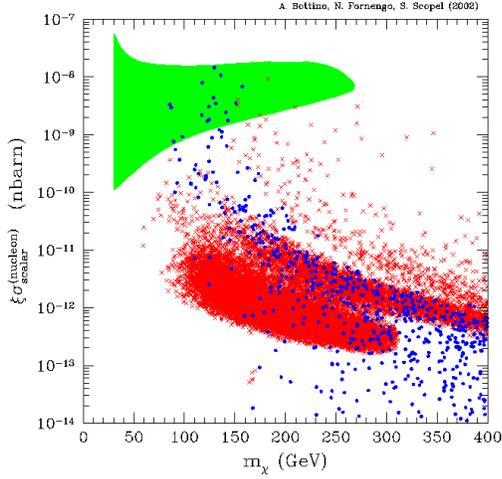}
\vspace{-50pt}
\caption{Scatter plot of the values of the product $\xi \sigma_{\rm
scalar}^{\rm (nucleon)}$ between the scaling parameter $\xi$ and the
neutralino--nucleon scalar cross section, vs. the neutralino mass
$m_\chi$, for a generic scan of the mSUGRA parameter space. (Red)
crosses refer to $0.05 \leq \Omega_\chi h^2 \leq 0.3$, while (blue)
dots denote configurations with $\Omega_\chi h^2 < 0.05$.  $m_b$,
$m_t$ and $\alpha_s$ are varied inside their 95\% C.L. allowed
ranges. The shaded (green) area \cite{distrib} shows the region which,
for a relic particle with pure coherent interactions, is compatible
with the annual modulation effect observed by the DAMA/NaI experiment
\cite{dama}.
\label{fig:sugra_sigma}}
\end{figure}

\begin{figure}[t] \centering
\vspace{-20pt}
  \includegraphics[width=19pc]{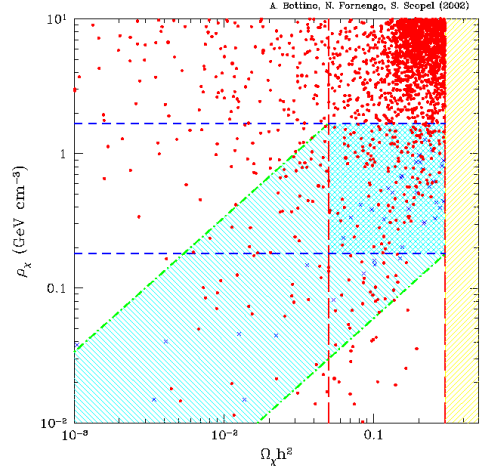}
\vspace{-50pt}
\caption{Scatter plot of the values of the neutralino local density
$\rho_\chi$ which are compatible with the current direct detection
sensitivity, vs. the neutralino relic abundance $\Omega_\chi h^2$, for
a generic scan of the mSUGRA parameter space. The horizontal (blue)
dashed lines delimit the allowed interval for the {\em total} dark
matter density, as obtained in the recent analysis of
Ref.\cite{distrib}. The vertical (red) long--dashed lines show the
range where the neutralino relic abundance lies in the cosmologically
relevant range. The slanted dashed band delimits the region where the
{\em neutralino} local dark matter density is reduced, with respect to
the nominal values for the {\em total} local dark matter density,
proportionally to the average density of neutralinos in the Universe.
\label{fig:sugra_rho_omega}}
\end{figure}

Neutralino relic abundance in mSUGRA models is shown in
Figs. \ref{fig:sugra_relic_tb30} and \ref{fig:sugra_relic_tb50}. All
the mSUGRA parameters are varied \cite{noi} except for $\tan\beta$
(the ratio of the two Higgs vev's) which has been fixed at two
representative values. Here and hereafter, $m_b$, $m_t$ and $\alpha_s$
are varied inside their 95\% C.L. allowed ranges. We notice that for
$\tan\beta \lsim 40$ the requirement that the neutralino relic
abundance does not conflict with the cosmological observation on the
amount of dark matter in the Universe poses severe constraints on the
mSUGRA parameter space. This is not the case for large values of
$\tan\beta$.

The results shown in Figs. \ref{fig:sugra_relic_tb30} and
\ref{fig:sugra_relic_tb50} are valid in a standard cosmological model
for the evolution of the Universe. The results may be quite different
in models of the early Universe where the temperature at the end of
the reheating phase is low \cite{lowreheating0}. In this
low--reheating models, the strong constraints which are induced on the
mSUGRA parameter space at $\tan\beta \lsim 40$ are strongly relaxed,
depending on the actual value of the reheating temperature
\cite{lowreheating}.  The requirement that neutralinos provide a
sizable contribution to dark matter, {\em i.e.} that the neutralino
relic abundance lies in the range $0.05 \lsim \Omega_{\rm M} h^2 \lsim
0.3$, implies that the reheating temperature of the Universe cannot be
smaller than about 1 GeV in mSUGRA models \cite{lowreheating}.

Direct detection relies on the scattering of dark matter particles off
the nuclei in a low--background detector. The detection rate depends
on the neutralino--nucleus scattering cross section, which is usually
dominated by the coherent interaction, and is sensitive to the local
properties of the neutralinos in the halo, {\em i.e.} its local
abundance $\rho_\chi$ and its local velocity distribution (for a
recent and exhaustive analysis on this topics, see
Ref. \cite{distrib}). The current sensitivity of direct detection
experiments on the neutralino--{\em nucleon} cross section is: ${\rm
few} \cdot 10^{-10} {\rm nbarn} \lsim \xi \sigma_{\rm scalar}^{\rm
(nucleon)} \lsim {\rm few} \cdot 10^{-8} {\rm nbarn}$ for neutralino
masses in the range: $30 {\rm ~GeV} \lsim m_\chi \lsim 300 {\rm ~GeV}$
\cite{distrib}. These ranges takes into account a large variability of
galactic halo models \cite{distrib}. The quantity $\xi\leq 1$ measures
the fraction of local dark matter to be ascribed to the neutralino
\cite{noi,distrib}: $\rho_\chi = \xi \rho_l$, where $0.18 \lsim
\rho_l/({\rm GeV}\;{\rm cm}^{-3}) \lsim 1.68$ \cite{distrib} denotes
the local value of the {\em total} halo dark matter.

Fig. \ref{fig:sugra_sigma} shows the results of our theoretical
calculations in the mSUGRA scheme. The quantity $\xi$, which
determines whether the neutralino is a dominant or sub--dominant dark
matter component, is calculated according to its relic abundance as:
$\xi = {\rm min}(1,\Omega_\chi h^2/0.05)$ \cite{noi}. The closed
shaded area \cite{distrib} represents the DAMA/NaI region which is
obtained when the annual modulation effect observed by the DAMA
Collaboration \cite{dama} is interpreted as due to a dark matter
particle whose interactions with nuclei are dominated by coherent
scattering. A careful and exhaustive modeling of the galactic halo
properties has been performed in obtaining the region showed in
Fig. \ref{fig:sugra_sigma} \cite{distrib}.

The question whether current direct detection sensitivities are
probing dominant or subdominant relic neutralinos may be answered in
terms of the plot shown if Fig.\ref{fig:sugra_rho_omega}, which
translates directly in terms of astrophysical and cosmological
quantities the direct detection results \cite{noi}. By considering the
current interval of sensitivities on the quantity $[ \rho_\chi \times
\sigma_{\rm scalar}^{\rm (nucleon)} ]$, the calculation of
$\sigma_{\rm scalar}^{\rm (nucleon)}$ allows us to determine the
values of $\rho_\chi$ which are required, for each susy configuration,
in order to provide a detectable signal
\cite{noi}. Fig.\ref{fig:sugra_rho_omega} shows the calculated values
of $\rho_\chi$ vs.  the neutralino relic abundance, for the mSUGRA
scheme. We see that a fraction of susy models overlap with the region
of main cosmological and astrophysical interest: $0.05 \lsim
\Omega_\chi h^2 \lsim 0.3$ and $0.18 \lsim \rho_\chi/({\rm GeV
cm}^{-3}) \lsim 1.68$. For points in this region, the neutralino is the
dominant component of dark matter both in the Universe and at the
galactic level. For points which fall inside the band delimited by the
slant dot--dashed lines, the neutralino would provide only a fraction
of the cold dark matter density both at the level of local density and
at the level of the average $\Omega$, a situation which would be
possible if the neutralino is not the unique cold dark matter particle
component. On the other hand, configurations above the upper
dot--dashed line and below the upper horizontal dashed line would
imply the somewhat more unlikely situation of a stronger clustering of
neutralinos in our halo as compared to their average distribution in
the Universe. Finally, configurations above the upper horizontal line
are incompatible with the upper limit on the local density of dark
matter in our Galaxy.

\section{Conclusions}

We can certainly define the following items as the current main issues
and open problems in particle dark matter studies: {\em i)} to explain
the observed amount of dark matter in the Universe ($0.05 \lsim
\Omega_{\rm M} h^2 \lsim 0.3$) by finding suitable particle
candidates; {\em ii)} to detect a relic particle. We have shown that
for both of these issues, there appear to be good prospects of
success, especially for the most studied candidate which is the
neutralino. In particular, there are many susy schemes where relic
neutralinos can provide enough cosmological abundance to explain the
observed amount of dark matter, and at the same time they can have
detection rates large enough to be accessible to detection. Clearly
the occurrence of this particularly interesting situation depends on
the actual realization of supersymmetry. The observation of a signal
from dark matter, like for instance in the case of the annual
modulation effect observed by the DAMA/NaI Collaboration or of signals
which could hopefully come in future experiments, can be very
important not only for astrophysics and cosmology but also for
particle physics, since the need to explain such effects can help in
deriving properties of particle physics models and possibly
discriminate among different realizations, for instance of
supersymmetry.


\begin{thebibliography}{9}

\bibitem{TAUP01_Proc} For a recent updated review on the subjects
concerning dark matter and dark energy, see the Proceedings of TAUP
2001, Nucl. Phys. (Proc.  Suppl.) B 110 (2002).

\bibitem{TAUP01} N. Fornengo, invited review
talk at TAUP 2001, Nucl. Phys. (Proc.  Suppl.) B 110 (2002) 26
[arXiv:hep-ph/0201156].

\bibitem{noi} A. Bottino, F. Donato, N. Fornengo and S. Scopel, Phys.  Lett.
  B423 (1998) 109 [arXiv:hep-ph/9709292]; Phys. Rev. D 59 (1999) 095003
  [arXiv:hep-ph/9808456]; Phys. Rev. D 59 (1999) 095004
  [arXiv:hep-ph/9808459]; Phys. Rev. D 62 (2000) 056006
  [arXiv:hep-ph/0001309]; Phys. Rev. D 63 (2001) 125003
  [arXiv:hep-ph/0010203].

\bibitem{others} For an updated list of references to theoretical
papers on direct detection, see Ref. \cite{TAUP01}.

\bibitem{lowreheating0} G.F. Giudice, E.W. Kolb and A. Riotto,
Phys. Rev. D64 (2001) 023508.

\bibitem{lowreheating} N. Fornengo, A. Riotto and S. Scopel, in
preparation.

\bibitem{distrib} P. Belli, R. Cerulli, N. Fornengo and S. Scopel,
arXiv:hep-ph/0203242, to appear in Phys. Rev. D.

\bibitem{dama} R. Bernabei {\em et al.} (DAMA/NaI Collaboration),
Phys.  Lett. B424 (1998) 195; Phys. Lett. B450 (1999) 448; Phys. Lett.
B480 (2000) 23.

\end{thebibliography}
\end{document}